\documentclass[preprint,prd,superscriptaddress,showpacs,
amssymb,amsmath,amsfonts,nofootinbib]{revtex4}

\usepackage{graphicx} 
\usepackage{latexsym}
\usepackage{dcolumn} 
\usepackage{bm} 

\def\NC   {\rm Nuovo Cimento }

\def\NPB {{\rm Nucl. Phys.} {\bf B}}
\def\PLB {{\rm Phys. Lett.} {\bf B}}

\def\PR  {{\rm Phys. Rev. }}

\def\PRD {{\rm Phys. Rev.} {\bf D}}

\def\ba{\begin{eqnarray}}\def\ea{\end{eqnarray}}
\def\bc{\begin{center}}\def\ec{\end{center}}
\def\nn{\nonumber\\}
\newcommand{\etal}{{\em et al.}}

\begin{document}

\title{Incoherent photoproduction of pseudoscalar mesons off nuclei at forward angles}

\author{S.~Gevorkyan}
\altaffiliation[Corresponding author:]{gevs@jlab.org.\\
On leave of absence from Yerevan Physics Institute
}
\affiliation{Joint Institute for Nuclear Research, Dubna, 141980, Russia}

\author{A.~Gasparian}
\affiliation{North Carolina A\&T State University, Greensboro, NC 27411, USA}

\author{L.~Gan}
\affiliation{University of North Carolina Wilmington, Wilmington, NC 28403, USA}

\author{I.~Larin}
\affiliation{Institute for Theoretical and Experimental Physics,  Moscow, Russia}

\author{M.~Khandaker}
\affiliation{Norfolk State University, Norfolk, VA 23504, USA}

\date{\today}

\begin{abstract}

Recent advances in the photon tagging facilities together with the novel,
high resolution fast calorimetry made possible to perform photoproduction
cross section measurements of pseudoscalar mesons on nuclei with a percent
level accuracy. The extraction of the radiative decay widths, needed for
testing the symmetry breaking effects in QCD, from these
measurements at small angles is done by the Primakoff method. This method
requires theoretical treatment of all processes participating in these
reactions at the same percent level. The most updated description of general
processes, including the nuclear coherent amplitude, is done in our previous
paper. In this work, based on the
framework of Glauber multiple scattering theory, we obtain analytical
expressions for the incoherent cross section of the photoproduction
of pseudoscalar mesons off nuclei accounting for the mesons'
absorption in nuclei and Pauli suppression at forward production
angles. As illustrations of the obtained formulas, we calculate the
incoherent cross section for photoproduction from a closed shell
nucleus, $^{16}$O, and from an unclosed shell nucleus, $^{12}$C.
These calculations allow one to compare different approaches and
estimate their impact on the incoherent cross section of the
processes under consideration.
\end{abstract}


\maketitle

\section{Introduction}
Recent years have seen a renewed interest in the study of
photoproduction of pseudoscalar mesons off complex nuclei
at few GeV energies due to the advent of high duty-cycle
electron accelerators together with new photon tagging
facilities and novel experimental technologies.  It has now
become possible to perform photoproduction cross section
measurements of pseudoscalar mesons off nuclei with a
percent level accuracy.

Photoproduction of pseudoscalar mesons off nuclei at small
angles,
\ba
\gamma+A \to Ps+A ~~~ (Ps = \pi^0, \eta, \eta^{\prime}),
\label{eq1}
\ea
is provided mainly by the possibility of extraction of
their radiative decay widths
from experimental data.  These reactions at
small angles proceed via production in the Coulomb field
of a heavy nucleus, the Primakoff effect~\cite{primakoff51}.
On the other hand, photoproduction
in the nuclear strong field complicates the extraction of
decay widths from experimental data and requires detail theoretical
examination.

The precision measurement of $\pi^0$ radiative decay width via
the Primakoff effect in the photoproduction of pseudoscalar
mesons off nuclear targets has recently been completed at
Jefferson Lab~\cite{gasparian07}.  The energy upgrade of the CEBAF
accelerator to 12 GeV will provide a great opportunity to extend
the current program to the $\eta$ and $\eta^{\prime}$ mesons.

In a recent article~\cite{gevorkyan09} we discussed that a
precise knowledge of the radiative decay widths of light
pseudoscalar mesons provide an accurate test of chiral anomalies
and mixing effects that are due to isospin breaking by the
difference of the masses of light quarks.  This is of great
importance to understand fundamental properties of QCD at
low energies~\cite{goity02,ananth02,ioffe07}.
The precision extraction of the decay widths requires state-of-the-art
theoretical descriptions of all processes participating in the reaction.
In our recent work~\cite{gevorkyan09}
we considered explicitly the photoproduction of pseudoscalar
mesons in the Coulomb field of nuclei and their coherent
photoproduction off the nuclear strong field.  In the present
work we will consider the incoherent photoproduction of
pseudoscalar mesons off nuclei, \emph{i.e.},
\ba
\gamma+A \to Ps+A^{\prime}.
\label{eq2}
\ea
Here $A^{\prime}$ is the target final state including all
possible nuclear excitations and its breakup.  As in our
previous work~\cite{gevorkyan09}, here we will discuss
the incoherent photoproduction of $\pi^0$ meson bearing
in mind that the obtained expressions, after obvious
modifications, can be applied to photoproduction of the
other pseudoscalar mesons.

The necessity for such investigations is two-fold.  As mentioned
above, the advent of precise experimental data demands the
accuracy of theoretical predictions to be at relevant levels.
Thus, the widely used expressions for the incoherent photoproduction
of pseudoscalar mesons~\cite{engelbrecht64,bellettini70}
cannot be considered as satisfactory.  The reason for this is that
the nucleon spin-nonflip amplitude\footnote{
The photoproduction of pseudoscalar mesons off the nucleon is
described by four independent amplitudes~\cite{chew57}.  In the
coherent photoproduction off spinless nuclei only the spin-nonflip
amplitude survives.  As for the incoherent photoproduction, this
amplitude is the dominant one~\cite{sibirtsev09}, though all four
amplitudes are involved in the process.
}
is zero at zero production angle due to conservation of angular
momentum.

As was first shown by F\"{a}ldt~\cite{faldt72}, such behavior of
the nucleon amplitude requires particular consideration for the
coherent photoproduction process off nuclei.  On the other hand,
in the incoherent production, as we will show later, similar
corrections to the cross section take place and their neglect
leads to meaningless results for the incoherent photoproduction
at forward angles.

Despite the fact that the incoherent production off nuclei in
non-diffractive processes was considered in the works of
F\"{a}ldt~\cite{faldt73,faldt77}, we are compelled to study
this process again since the expressions obtained in
Refs.~\cite{faldt73,faldt77} are not suitable for the production
at small angles\footnote{
For instance, the incoherent cross section would be strongly
suppressed as a result of Pauli exclusion principle, which is
not the case in Ref.~\cite{faldt77} (see, for instance, expression (33b)).
}.
Pauli suppression is at work for production at small angles and
its correct accounting is crucial for the process under
consideration.

\section{The incoherent photoproduction off nuclei}
In the Glauber multiple scattering theory~\cite{glauber67} the
amplitude of the process as a result of which the nuclear wave
function in the initial state $\Phi_i(\vec r_1,\vec r_2 ...\vec r_A)$
transforms to the wave function
of the final state $\Phi_f(\vec r_1,\vec r_2 ...\vec r_A)$ is
\ba
F_{fi}(\vec q,\Delta) &=& \frac{ik}{2\pi}
\int d^2be^{i\vec q\cdot\vec b}d^3r_1...d^3r_A
\Phi^*_f(\vec r_1,\vec r_2 ...\vec r_A) \nn
&\times& \sum_{j=1}^{A}
\Gamma_j(\vec b,\vec r_1...r_A)
\Phi_i(\vec r_1,\vec r_2 ...\vec r_A), \nn
\Gamma_j(\vec b,\vec r_1...\vec r_A) &=& \Gamma_p(\vec b-\vec s_j)
e^{i\Delta z_j} \nn
&\times& \prod_{i\neq j}^A
\big[1-\Gamma_s(\vec b-\vec s_i)\theta(z_i-z_j)\big].
\label{eq3}
\ea
Here the two-dimensional vectors $\vec b$ and $\vec s_i$ are the
impact parameter and the nucleon's transverse coordinate, respectively;
$z_i$ is the nucleon's longitudinal coordinate in the nucleus;
$\vec q$ and $\Delta$ are the transverse and longitudinal components of
the transferred momenta:
$\displaystyle q^2=4kp \sin^2\left(\theta/2\right)$,
$\displaystyle \Delta=M^2/(2k)$.
Here $k$ and $p$ are the photon and meson momenta and $M$ is the
meson mass.  The profile functions $\Gamma_{p,s}(\vec b-\vec s)$, by
definition, are the two-dimensional Fourier transformations of the
elementary amplitudes for the photoproduction of pseudoscalar meson
off the nucleon, $f_p=f(\gamma+N\to Ps +N)$, and elastic meson-nucleon
scattering, $f_s=f(Ps+N\to Ps+N)$:
\ba
\Gamma_{p,s}(\vec b-\vec s) = \frac{1}{2\pi ik}\int e^{i\vec q\cdot
(\vec b-\vec s)}f_{p,s}(q)d^2q.
\label{eq4}
\ea
The summed cross section, including both the coherent and the
incoherent production processes, can be obtained if one uses the
closure approximation, \emph{i.e.}, assumes that in the limits of
the produced particle energies (usually 50-100 MeV) the final states
of the nucleus forms a complete set:
\ba
\sum_{n}\Phi_f\Phi_f^* = 1.
\label{eq5}
\ea
Under this assumption the summed cross section depends only on the
ground state wave function~\cite{faldt73}:
\ba
\frac{d\sigma}{d\Omega} &=& \sum_f \left|F_{if}\right|^2 \nn
&=& \frac{k^2}{(2\pi)^2}\int e^{i\vec q\cdot(\vec b-\vec b')}d^2bd^2b'
d^3r_1...d^3r_A \left| \Phi_i(\vec r_1...\vec r_A) \right|^2 \nn
&\times& \sum_{j,j'} \Gamma_j(\vec b,\vec r_1...\vec r_A,\Delta)
\Gamma_{j'}^*(\vec b',\vec r_1...\vec r_A,\Delta).
\label{eq6}
\ea

Let us assume that the ground state of the nucleus can be
described by means of the independent particle model.  Introducing the
single particle density $\rho(r)$ we have
\ba
\mid \Phi_i(\vec r_1, ...\vec r_A)\mid^2 ~=
\prod_{k=1}^{A}{\frac{\rho(r_k)}{A}};~~~~\int \rho(\vec r) d^3r=A.
\label{eq7}
\ea
Separating the contributions of diagonal and non-diagonal terms in
the sum in Eq.~\ref{eq6}, one gets~\cite{faldt73,gevorkyan72}:
\ba
\frac{d\sigma_d}{d\Omega} &=& \frac{k^2}{(2\pi)^2}\int e^{i\vec q\cdot
(\vec b-\vec b')}d^2bd^2b'd^2sdz \nn
&\times& \Gamma_p(\vec b-\vec s)\Gamma_p^*(\vec b'-\vec s)
\rho(\vec s,z)e^{-E(\vec b,\vec b',z)}, \nn
E(\vec b,\vec b',z) &=& \int_{z}^{\infty} \big[\Gamma_s(\vec b-\vec s)+
\Gamma_s^*(\vec b'-\vec s) \nn
&-&  \Gamma_s(\vec b-\vec s)\Gamma^*_s(\vec b'-\vec s)\big]
\rho(\vec s, z')d^2sdz', \nn
\frac{d\sigma_{nd}}{d\Omega}&=&\frac{(A-1)}{A}\frac{k^2}{(2\pi)^2} \nn
&\times&
2 \Re \int e^{i\vec q\cdot(\vec b-\vec b')+i\Delta(z_2-z_1)} \nn
&\times& \theta (z_2-z_1)d^2bd^2b' d^2s_1d^2s_2 dz_1dz_2 \nn
&\times& \Gamma_p(\vec b-\vec s_1)
\Gamma_p^*(\vec b'-\vec s_2)\big[1-\Gamma_s(\vec b'-\vec s_2)\big] \nn
&\times&
\rho(\vec s_1,z_1)\rho(\vec s_2,z_2)e^{-E(\vec b,\vec b', z_1,z_2)}, \nn
E(\vec b,\vec b',z_1,z_2)
&=& \int_{z_1}^{z_2}\Gamma_s(\vec b-\vec s )\rho(\vec s,\vec z )d^2s dz \nn
&+& \int_{z_2}^{\infty}\big[\Gamma_s(\vec b-\vec s)
+ \Gamma_s^*(\vec b'-\vec s) \nn
&-&  \Gamma_s(\vec b-\vec s)\Gamma^*_s(\vec b'-\vec s)\big]
\rho(\vec s,\vec z)d^2sdz.
\label{eq8}
\ea
These expressions are general and can be applied to diffractive
production off nuclei (for instance, in photoproduction of vector
mesons $V(\rho,\omega,\phi,\psi)$) and to non-diffractive processes
like production of pseudoscalar mesons.

The incoherent processes at large momentum transfer were considered
in detail in Ref.~\cite{faldt73}\footnote{
In Refs.~\cite{faldt72,faldt73} the final formulas have been cited
for the production processes with the same absorption in nuclei
for the initial and final hadrons, which significantly facilitates
the calculations, but it is not appropriate for photoproduction.
}.
Our main goal here is to investigate the incoherent photoproduction of
pseudoscalar mesons at small momentum transfers.  In this kinematic
region one can neglect the mesons' multiple scattering to non-forward
angles, the effect important at high momentum transfers~\cite{glauber70}.
In this approximation the differential cross section Eq.~\ref{eq8} takes
the form:
\ba
\frac{d\sigma}{d\Omega}&=&
\frac{d\sigma_d}{d\Omega}+\frac{d\sigma_{nd}}{d\Omega} \nn
&=&\int{\rho(\vec s,z)|h(\vec s,z)|^2d^2sdz} \nn
&+& \frac{(A-1)}{A}\left|\int \rho(\vec s,z)h(\vec s,z)d^2sdz\right|^2, \nn
h(\vec s,z) &=& \frac{ik}{2\pi}\int e^{i\vec q\cdot\vec b+i\Delta z}
d^2b \Gamma_p(\vec b-\vec s) \nn
&\times& \exp{\Big(-\int_{z}^{\infty}
\Gamma_s(\vec b-\vec s')\rho(\vec s',z')d^2s'dz'\Big)}.
\label{eq9}
\ea
The second term in this expression includes the coherent cross section.
To obtain the incoherent cross section one has to subtract it from the
summed cross section
(Eq.~\ref{eq9}):
\ba \frac{d\sigma_{inc}}{d\Omega} &=&
\int{\rho(\vec s,z)|h(\vec s,z)|^2d^2sdz} \nn
&-& \frac{1}{A}\left |\int{\rho(\vec s,z)h(\vec s,z)d^2sdz)}\right |^2.
\label{eq10}
\ea
The well known approximation that is valid for intermediate and heavy
nuclei for convolution of the profile function $\Gamma_s(\vec b-\vec s)$
with nuclear density~\cite{glauber70} is
\ba
\int\Gamma_s(\vec b-\vec s)\rho(s,z)d^2sdz=\frac{\sigma'}{2}
\int\rho(\vec b,z)dz,
\label{eq11}
\ea
where $\displaystyle \sigma'= \frac{4\pi}{ik}f_s(0)$.
On general grounds it is known that the spin-nonflip nucleon amplitude
in photoproduction of pseudoscalar mesons would vanish at zero angles.
Thus, we choose for the elementary photoproduction amplitude on the
nucleon the parametrization
\ba
f_p(q)=(\vec n\cdot\vec q )\varphi (q).
\label{eq12}
\ea
Here $\displaystyle \varphi(0)\neq 0;
~\vec n=\frac{\vec k\times\vec\epsilon}{k}$,
with $\epsilon$ the polarization vector of the photon. The relevant cross
section for photoproduction off the nucleon is
$\displaystyle \frac{d\sigma_0}{d\Omega}=\frac{1}{2}q^2|\varphi (q)|^2$,
and the relevant profile function becomes
\ba
\Gamma_p(\vec b-\vec s)&=&-i\vec n\cdot
\frac{\partial\Gamma(\vec b-\vec s)} {\partial\vec b}, \nn
\Gamma(\vec b-\vec s)&=&\frac{1}{2\pi ik}
\int e^{i\vec q\cdot(\vec b-\vec s)}\varphi(q)d^2q.
\label{eq13}
\ea
Accounting for the fact that the profile function
$\Gamma_p(\vec b-\vec s)$ changes much slower than the nuclear density
$\rho(\vec s,z)$, it is straight forward to obtain from
Eqs.~\ref{eq9} and~\ref{eq13}
\ba
h(\vec s,z)
&=& \left((\vec n\cdot\vec q)-i\frac{\sigma'}{2}\vec n\cdot
\frac{\partial T(\vec s,z)}{\partial\vec s}\right)
\varphi(q)e^{i\vec q\cdot\vec s+i\Delta z} \nn
&\times& \exp{\Big(-\frac{\sigma'}{2}T(s,z)\Big)}, \nn
T(s,z)&=&\int_{z}^{\infty}\rho(\vec s,z')dz'.
\label{eq14}
\ea
Substituting this expression into Eq.~\ref{eq10}, we get for the
incoherent cross section of the process under consideration:
\ba
\frac{d\sigma_{inc}}{d\Omega} &=& \frac{d\sigma_0}{d\Omega}(q)
\left (N(0,\sigma)-\frac{|F(q,\Delta)|^2}{A}\right) \nn
&+& |\varphi(q)|^2 ~\frac{\sigma^2}{8}
\int \rho(\vec s,z) \left| \frac{\partial T(\vec s,z)}{\partial s}\right|^2 \nn
&\times& \exp{\Big(-\sigma T(\vec s,z)\Big)}d^2sdz, \nn
F(q,\Delta) &=& \int e^{i\vec q\cdot\vec s+i\Delta z}\rho(\vec s,z)
d^2sdz\exp{\Big(-\frac{\sigma'}{2}T(\vec s,z)\Big)} \nn
&-& \frac{\pi\sigma'}{q}\int \hspace{-0.10cm}J_1(qs)\rho(\vec s,z_2)
\frac{\partial\rho(\vec s,z_1)}
{\partial s} s ds dz_1dz_2 e^{i\Delta z_1} \nn
&\times& \exp{\Big(-\frac{\sigma'}{2}T(s,z_1)\Big)}, \nn
N(0,\sigma) &=& \int \frac{1 -
\exp{\left(-\sigma \int \rho(s,z)dz\right)}}{\sigma}d^2s.
\label{eq15}
\ea
Let us mention the main difference between this expression and the one
widely used for the incoherent cross
section~\cite{engelbrecht64,bellettini70}:
\ba
\frac{d\sigma_{inc}}{d\Omega}&=&\frac{d\sigma_0}{d\Omega}(q)
N(0,\sigma)\left [1- G(t)\right].
\label{eq16}
\ea
At small momentum transfer this expression goes to zero
($G(0)=1$).  On the other hand, this is not the
case for Eq.~\ref{eq15}, which differs from zero at $q=0$.  The
complete suppression takes place only if one neglects the final
state interaction of the produced meson ($\sigma=0$).
This fact has been well known for many years for incoherent cross section
of elastic scattering~\cite{glauber70} and for diffractive
photoproduction~\cite{moniz69,moniz71}.

\section{Pauli correlations}
Up to now we have assumed the factorization for the nuclear ground
state wave function (see Eq.~\ref{eq7}). As has been shown above,
even in such simplified model the orthogonality and completeness of
final states nuclear wave functions (closure approximation) lead
to suppression of the pseudoscalar mesons' yield at forward angles
different from the common one due to the mesons' final state
interactions in nuclei.

Let us estimate the impact of the nucleon-nucleon correlations on
the incoherent cross section.  Here we are constrained by the
correlations due to the Pauli exclusion principle, as their contribution
is the largest one and they determine the behavior of the cross
section at small momentum transfers.

The ground state wave function of the nucleus can be cast in the
form~\cite{moniz71}
\ba
\rho^{(A)}(\vec r_1,\vec r_2...\vec r_A)
&=& \hspace{-0.05cm}|\Phi_i(\vec r_1,\vec r_2...\vec r_A)|^2
= \hspace{-0.05cm}\rho(\vec r_1)\rho(\vec r_2)...\rho(\vec r_A) \nn
&+&
\hspace{-0.25cm}\sum_{{\rm contraction}}{\hspace{-0.25cm}\big[\rho(\vec r_1)\rho(\vec r_2)...
\rho(\vec r_A)\big]}+...,
\label{eq17}
\ea
where the pair contraction is defined by
$\Delta(\vec r_1,\vec r_2)=\rho^{(2)}(\vec r_1,\vec r_2)-\rho(\vec r_1)
\rho(\vec r_2)$, and the single-particle and two-particle densities
\ba
\rho(\vec r_1)
\hspace{-0.20cm} &=& \hspace{-0.20cm} \int \rho^{(A)}(\vec r_1,\vec r_2...\vec r_A)
d^3r_2d^3r_3...d^3r_A, \nn
\rho^{(2)}(\vec r_1,\vec r_2)
\hspace{-0.20cm} &=& \hspace{-0.20cm} \int \rho^{(A)}(\vec r_1,\vec r_2...\vec r_A)
d^3r_3d^3r_4...d^3r_A.
\label{eq18}
\ea
Let us introduce the so called mixed density~\cite{engelbrecht64}
\ba
d(\vec r_1,\vec r_2)=\sum_{j=1}^{A}\phi_j(\vec r_1)\phi_j^*(\vec r_2),
\label{eq19}
\ea
where $\phi_j(\vec r)$ is the wave function of the single nucleon, so the
nuclear density is given by $\rho(\vec r)=d(\vec r,\vec r)$.
The two-body density can be expressed through the mixed density
\ba
\rho^{(2)}(\vec r_1,\vec r_2)&=&\frac{1}{A(A-1)}\big [d(\vec r_1,\vec r_1)
d(\vec r_2,\vec r_2) \nn
&-& d(\vec r_1,\vec r_2)d(\vec r_2,\vec r_1)\big].
\label{eq20}
\ea
Substituting this expression into the general formula, Eq.~\ref{eq6}, for
the summed cross section and using the relation
\ba
\int d(\vec r_1,\vec r_2)d(\vec r_2,\vec r_1)d^3r_2=d(\vec r_1,\vec r_1),
\label{eq21}
\ea
we obtain
\ba
\frac{d\sigma_{inc}}{d\Omega}&=&\int d(\vec s,z,\vec s,z)
|h(\vec s,z)|^2d^2sdz \nn
&-& \int d(\vec s_1,z_1,\vec s_2,z_2)d^*(\vec s_2,z_2,\vec s_1,z_1) \nn
&\times& h(\vec s_1,z_1)h^*(\vec s_2,z_2)d^2s_1dz_1d^2s_2dz_2.
\label{eq22}
\ea
It is straight forward to check that in the Born approximation
($\sigma=0$) this expression takes the well known
form~\cite{engelbrecht64}
\ba
\frac{d\sigma_{inc}}{d\Omega}&=& A\frac{d\sigma_0}{d\Omega}(1-G(q, \Delta)); \nn
G(q,\Delta)&=&\int e^{i\vec q(\vec s_1-\vec s_2)+i\Delta (z_1-z_2)} \nn
&\times& d(\vec s_1,z_1,\vec s_2,z_2) d^*(\vec s_2,z_2,\vec s_1,z_1) \nn
&\times& d^2s_1d^2s_2dz_1dz_2.
\label{eq23}
\ea

To proceed with the calculations for the incoherent cross section, one
has to choose the model for the mixed density.  We will consider the
photoproduction  from light nuclei, for which the shell model with
harmonic-oscillator wave functions works well.

Let us first consider a closed shell nucleus, for instance $^{16}$O.
The mixed density in the independent particle model for this nucleus
with harmonic-oscillator wave functions~\cite{engelbrecht64} is
\ba
d(\vec r_1,\vec r_2)=\frac{4}{a_0^3\pi^{3/2}}
\left(1+2\frac{\vec r_1\vec r_2}{a_0^2}\right)
\exp{\left(-\frac{r_1^2+r_2^2}{2a_0^2}\right)}.
\label{eq24}
\ea
Substituting this equation into Eq.~\ref{eq22} we calculate the
incoherent cross section.

\begin{figure}[!htp]
\begin{center}
\includegraphics [scale=0.35]{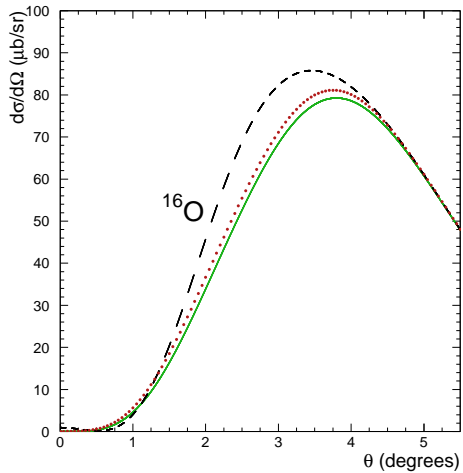}
\caption{
Incoherent differential cross section for $\pi^0$ photoproduction off
$^{16}$O nucleus as a function of production angle in the lab system.
The long-dashed line is calculated by Eq.~\ref{eq15} with single-nucleon
density relevant to harmonic-oscillator wave functions; the dotted line
is the incoherent cross section taking into account Pauli correlations,
Eqs.~\ref{eq22} and~\ref{eq24}; the solid line is the standard
parametrization as in Eq.~\ref{eq16}.
}
\label{fig:inc_16O}
\end{center}
\end{figure}

In Fig.~\ref{fig:inc_16O} we plot the dependence of the incoherent
cross section for photoproduction of $\pi^0$ mesons off $^{16}$O
nucleus as a function of the production angle.  The oscillator parameter
for $^{16}$O is $a_0=1.81 ~{\rm fm}$~\cite{hofstadter67}, and the
total cross section of $\pi^0N$ interaction is $\sigma=27 ~{\rm mb}$.
The solid line is calculated by Eq.~\ref{eq16}.  The long-dashed and
dotted curves are the results of calculations according to
Eqs.~\ref{eq15} and~\ref{eq22}, respectively.

For the $^{12}$C nucleus, we choose the mixed density in the matrix
form~\cite{tarasov76}:
\ba
d(\vec r_1,\vec r_2)&=&\frac{2}{a_0^3\pi^{3/2}}\big[1+\frac{2}{3a_0^2}
\left(2\vec r_1\cdot\vec r_2+i\vec\sigma
\cdot(\vec r_1\times\vec r_2)\right)\big] \nn
&\times& \exp{\left(-\frac{r_1^2+r_2^2}{2a_0^2}\right)}.
\label{eq25}
\ea
In this case the single-particle nucleon density is expressed through
mixed density by the relation
\ba
\rho(\vec r)={\rm Tr} ( d(\vec r,\vec r)).
\label{eq26}
\ea

\begin{figure}[!htp]
\begin{center}
\includegraphics [scale=0.35]{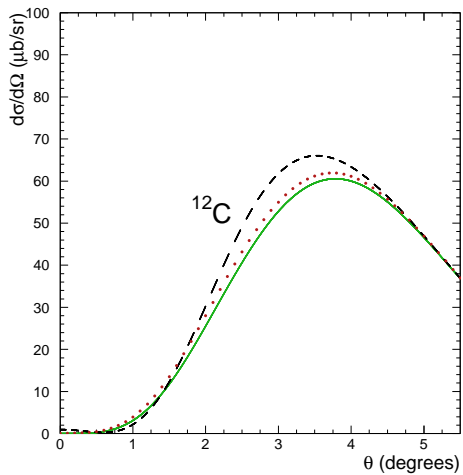}
\caption{
Incoherent differential cross section for $\pi^0$ photoproduction off
$^{12}$C nucleus as a function of the production angle in the lab system.
The curves have the same meaning as in Fig.~\ref{fig:inc_16O}.
}
\label{fig:inc_12C}
\end{center}
\end{figure}

The results of our calculations for $^{12}$C using mixed density,
Eq.~\ref{eq25}, with oscillator parameter
$a_0=1.6 ~{\rm fm}$~\cite{glauber70} are shown in Fig.~\ref{fig:inc_12C}.
The solid line is calculated using Eq.~\ref{eq16}.
The long-dashed and the dotted curves have the same meaning as in
Fig.~\ref{fig:inc_16O}.

As seen from the figures, the correct accounting of pion absorption
changes the behavior of the incoherent cross sections at forward
angles which can be crucial for the extraction of
radiative decay widths of pseudoscalar mesons.

\section{Summary and conclusions}
In recent years it has become possible to perform high precision
measurements of photoproduction cross sections for light pseudoscalar
mesons on nuclei.  The extraction
of meson radiative decay widths with high
accuracy from Primakoff type of experiments requires state-of-the-art
theoretical descriptions of all participating processes.
We have calculated the incoherent photoproduction cross sections for $\pi^0$ 
using the Glauber multiple scattering theory.
On that way, we have revised the existing approaches for
incoherent non-diffractive photoproduction at forward angles and
obtained new expressions (see Eq.~\ref{eq15}) which correctly account
for meson absorption in the final state.
This expression correctly takes into account the fact that the differential cross
section on a nucleon is zero at zero production angle.  Moreover, the
zero in the nucleon cross section leads to additional terms in the
incoherent cross section without which the incoherent cross section
would be negative at very small angles.

Pauli correlations between nucleons on which photoproduction takes place
are properly taken into account as their effects are dominant at forward
angles.  To account for Pauli suppression, the ground state wave functions
of nuclei have to be antisymmetrized as the nucleons are fermions.  This
can be done if one works using the mixed densities, as in~\cite{engelbrecht64}.

Taking into account the above considerations, the general expressions of
the incoherent photoproduction
cross sections for pseudoscalar mesons off nuclei were obtained for the first
time (see Eq.~\ref{eq22}).
Using these new expressions and taking the mixed and single nucleon
densities in correspondence with the harmonic-oscillator model, we have
performed calculations for a closed shell light nucleus (oxygen) and
an unclosed shell nucleus (carbon).

\section{Acknowledgments}
This work was inspired by the PrimEx experiment at Jefferson Lab to
measure the $\pi^0$
radiative decay width via the Primakoff effect with high precision.
It was partially supported by the National Science Foundation grants
PHY-0245407 and PHY-0555524.  One of the authors (S.G.) is grateful to
Alexander Tarasov for useful comments.

\end{document}